\documentclass[aps,prx,reprint]{revtex4-1}

\usepackage{graphicx}
\usepackage{mathtools,amsthm,amssymb,braket}

\usepackage[utf8]{inputenc}
\usepackage[german,  danish, english]{babel}

\usepackage[pdfauthor={V. O. Shkolnikov, Roman Mauch and Guido Burkard},
pdftitle={All-microwave holonomic control of an electron-nuclear two-qubit register in diamond},
pdfsubject={},colorlinks=true,linkcolor=blue,citecolor=magenta]{hyperref}

\usepackage{bbm}

\begin{document}

%%%%%%%%%%%%%%%%%%%%%%%%%%%%%%%%%%%%%%%%%%%%%%%%%%%%%%%%%%%%%%%

\title{All-microwave holonomic control of an electron-nuclear two-qubit register in diamond}

\author{V. O. Shkolnikov}
\author{Roman Mauch}
\author{Guido Burkard}
\affiliation{Department of Physics, University of Konstanz, D-78457 Konstanz, Germany}

%\date{\today}

%%%%%%%%%%%%%%%%%%%%%%%%%%%%%%%%%%%%%%%%%%%%%%%%%%%%%%%%%%%%%%%
%%% Abstract
%%%%%%%%%%%%%%%%%%%%%%%%%%%%%%%%%%%%%%%%%%%%%%%%%%%%%%%%%%%%%%%

\begin{abstract}
	We present a theoretical scheme that allows to perform a universal set of holonomic gates on a two qubit register, formed by a $^{13}$C nuclear spin coupled to the electron spin of a nitrogen-vacancy center in diamond. Strong hyperfine interaction between the electron spin and the spins of the first three shells of $^{13}$C atoms allows to operate the state of the register on the submicrosecond timescale using microwave pulses only. We describe the system and the operating regime analytically and numerically, as well as simulate the initialization protocols.
\end{abstract}

\maketitle
%%%%%%%%%%%%%%%%%%%%%%%%%%%%%%%%%%%%%%%%%%%%%%%%%%%%%%%%%%%%%%%
%%% Bibliography
%%%%%%%%%%%%%%%%%%%%%%%%%%%%%%%%%%%%%%%%%%%%%%%%%%%%%%%%%%%%%%%
\section{Introduction}
The negatively charged nitrogen vacancy (NV$^-$) center in diamond is a point defect, that has attracted significant attention in the recent years. Its bright optical transition with the zero phonon line of $1.945\text{ eV}$ \cite{1367-2630-13-2-025025,DOHERTY20131} and the existence of intersystem crossing provides a good mechanism for initialization and read out of a defect's spin state \cite{PhysRevLett.92.076401}. The ground state of the NV$^-$ center is a spin triplet and is sensitive to magnetic and electric fields, as well as to strain \cite{doi:10.1021/acs.nanolett.6b04544, PhysRevB.85.205203, PhysRevB.98.075201}, which makes the defect useful for metrological applications \cite{Balasubramanian2008, Maze2008, Barfuss2015}. The long lifetime of the ground state coherence \cite{PhysRevB.80.041201} together with the fast optical initialization and readout renders the NV center interesting for quantum information purposes \cite{Hensen2015,Togan2010}. Recently fault tolerant universal geometric single-qubit gates \cite{SJOQVIST201665} have been achieved both using optical \cite{PhysRevLett.119.140503,Sekiguchi2017} and microwave \cite{Zu2014,Arroyo-Camejo2014} control of the spin. Scaling up to many NV spins is still an issue, as it is challenging to couple the defect spins \cite{Dolde2013, Pfaff2012, PhysRevB.95.205420}. On the other hand, numerous experiments have been performed on multiqubit registers that include the NV center electron spin coupled to the nearby $^{13}$C nuclear spins through the hyperfine interaction \cite{{Childress281,Dutt1312,Neumann1326,Waldherr2014,Yang2016,Maurer1283}}. Such a configuration allows the use of the longer coherence time of the nuclear spin to preserve the quantum state during times exceeding the $T_2^*$ of the electron spin. This can then be used for distributed quantum computation with electron-nuclear quantum registers \cite{PhysRevA.76.062323} or to gain increased sensitivity in metrological applications of NV centers \cite{Zaiser2016}. From this perspective the feasibility of universal control of the state of such registers becomes important. The existing experiments described in the literature \cite{PhysRevLett.93.130501} allow fast microwave control of the electron spin, as well as fast entangling CNOT or CPHASE gates controlled with the state of the nuclear spin. At the same time performing single qubit gates on the nuclear spins that are relatively close to the electron spin still required radio frequency pulses, that weakly couple to the nuclear spins due to their low gyromagnetic ratio \cite{PhysRevLett.102.210502}. 

Our work is motivated by the fact that the hyperfine interaction between the nearest neighbour $^{13}$C nuclear spin and an NV center provides a nuclear spin splitting of the order of $130\text{ MHz}$ \cite{PhysRevB.80.241204, PhysRevB.79.075203}, which allows for universal holonomic \cite{SJOQVIST201665} single and two-qubit gates on the two-qubit register, assisted by hyperfine interaction. The key enabling idea is to use a magnetic field to mix the electronic states $\ket{-1}$ and $\ket{0}$. In this case the quantization axis for the nuclear spin will depend on the state of the electron spin. We will show that this implies that electronic transitions between different hyperfine levels are no longer forbidden by nuclear spin selection rules and can efficiently be driven by microwaves. This should result in a speed-up compared to the existing schemes and provide universal control of the register, requiring application of microwave-only pulses and making use of the relatively stronger electron magnetic dipolar transitions.

This paper is structured as follows. In section \ref{Sec2} we will consider our scheme in the leading order of perturbation theory, providing a more detailed treatment in the Appendix \ref{appA}. In section \ref{Sec3} we will discuss how one could initialize and read out the state of the two-qubit register. Section \ref{Sec4} will be concerned with the construction of the pulses for universal quantum computing on the two-qubit system.

\section{System and operating regime}
 \label{Sec2}
 The Hamiltonian describing the ground state of the NV interacting with the nuclear spin of a nearby $^{13}$C is
 \begin{eqnarray}
 \begin{aligned}
 &\hat{H}_{gs}=\hat{H}_e+\hat{H}_n+\hat{H}_{hf},\\
 &\hat{H}_e=D_{gs}\hat{S}_z^2+\gamma_e\mathbf{B}\cdot\mathbf{\hat{S}},\\
 &\hat{H}_n=\gamma_n\mathbf{B}\cdot\mathbf{\hat{I}},\\
 &\hat{H}_{hf}=\sum_{i,j=\{x,y,z\}}A_{ij}\hat{S}_i\hat{I}_j.
 \label{hyperfine_Hamiltonian}
 \end{aligned}
 \end{eqnarray}
 Here $D_{gs}=2.88$ GHz is the ground state zero-field splitting, $\gamma_e=g\mu_B=2.8$ MHz/G is the electronic gyromagnetic ratio and $\gamma_n=0.001$ MHz/G is the nuclear gyromagnetic ratio. The values for the hyperfine tensor $A_{ij}$ are taken from \cite{PhysRevB.80.241204}. This tensor is approximately diagonal in the basis, where the z-axis coincides with the direction connecting the vacancy to the $^{13}$C atom. The eigenvalue corresponding to this axis amounts to $201$ MHz for a $^{13}$C atom in the first coordination shell, while the other two eigenvalues are $120$ MHz. In our simulations we do a basis change to obtain the values of the tensor in the NV-axial basis. In what follows we will neglect the splittings arising due to the Zeeman Hamiltonian $\hat{H}_n$, as they are much smaller than those arising from hyperfine interaction for the $^{13}$C atom in the first coordination shell of an NV center.

\begin{figure}[hb]
	\includegraphics[width=0.4\textwidth]{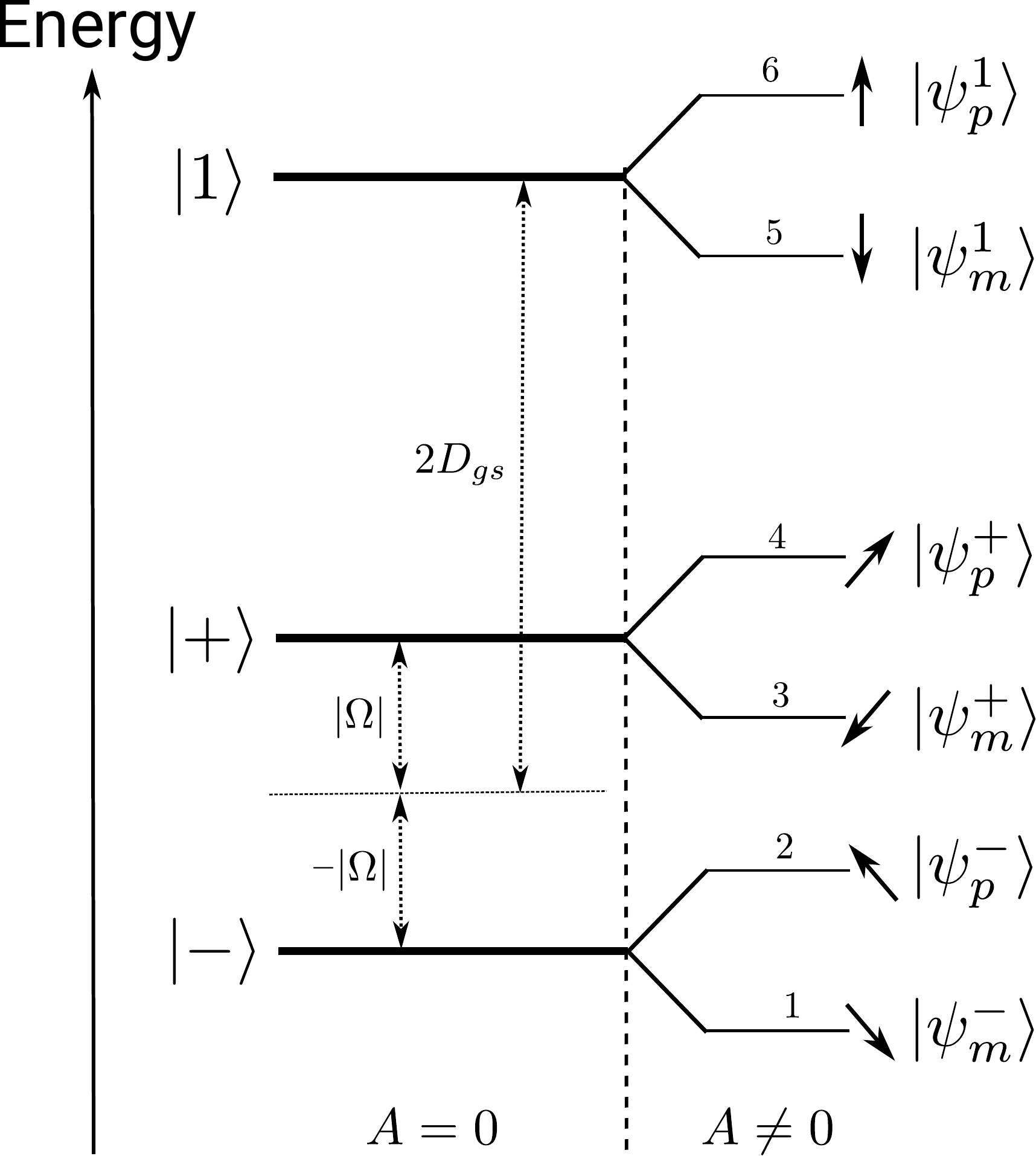}
	\caption{Energy level structure of the ground state of the NV$^{-}$ center in diamond coupled to a nearby $^{13}$C nuclear spin when a nonparallel magnetic field mixes the electronic levels $\ket{0}$ and $\ket{-1}$. The states $\ket{+}$, $\ket{-}$ and $\ket{1}$ to the left of the dashed line are the eigenstates of the electron spin without hyperfine interaction (hyperfine tensor $A=0$), the direction of the black arrows to the right of the dashed line indicates the quantization axis of the nuclear spin and $\ket{\psi_{p,m}^e}$ ($e\in\{+,-,1\}$) mark the corresponding hyperfine eigenstates. $D_{gs}$ is the NV$^{-}$ ground state zero field splitting, while $|\Omega|$ marks the splitting that arises between $\ket{+}$ and $\ket{-}$ when nonparallel magnetic field mixes $\ket{0}$ and $\ket{-1}$.}
	\label{fig:levels}
\end{figure}

Let us first treat the electronic part of the Hamiltonian (\ref{hyperfine_Hamiltonian}), later we will add the hyperfine interaction as a perturbation.
We first assume magnetic field $B_z=D_{gs}/\gamma_e$ in the direction of the NV symmetry axis, so that the levels $\ket{-1}$ and $\ket{0}$ are degenerate. Now we also assume a magnetic field $B_\perp$ in the direction perpendicular to the symmetry axis, and write the magnetic field as $B_x-iB_y=B_\perp e^{i\phi}$. The coupling of $\ket{1}$ to $\ket{0}$ is suppressed due to a large energy gap $2D_{gs}$ between them, therefore, in this section we neglect this coupling. Thus the eigenvectors of the Hamiltonian will be $\ket{+}=\left(e^{i\phi/2}\ket{0}+e^{-i\phi/2}\ket{-1}\right)/\sqrt{2}$ and $\ket{-}=\left(e^{i\phi/2}\ket{0}-e^{-i\phi/2}\ket{-1}\right)/\sqrt{2}$ with the energies $\pm|\Omega|$ respectively, with $|\Omega|=\gamma_eB_\perp/\sqrt{2}$. Now we add the hyperfine interaction $\hat{H}_{hf}$ to the system. Assuming $||A||\ll|\Omega|$, we can restrict our analysis to the secular terms of the hyperfine interaction, that do not flip the electron spin. Then for each electronic level we can describe the hyperfine interaction in terms of the Knight field $h_j$, acting on the nuclear spin $I_j$,
\begin{equation}
\begin{split}
	&\hat{\tilde{H}}_{hf}=\sum_{e=\{+,-,1\}}\sum_{j=\{x,y,z\}}\ket{e}\bra{e}h_j^{e}\hat{I}_j,\\
	&h_j^{e}=\sum_{i=\{x,y,z\}}A_{ij}\braket{e|\hat{S}_i|e}.
\end{split}
\end{equation}
 One of the key ingredients of the current proposal is the fact that the Knight field turns out to point in different directions for each of the three electronic levels, resulting in the level structure of the defect's ground state shown in Fig. \ref{fig:levels}. 
We can thus conclude that transitions between all eigenlevels of the system are now allowed by the nuclear spin selection rules and can be driven with a microwave field, oriented in the direction perpendicular to the symmetry axis of the defect. Let us number the levels in the figure from bottom to the top with $1,...,6$. For a given direction of the microwave field, resonant with the transition from level $i$ to level $j$, the Hamiltonian of the microwave field takes the form
 \begin{equation}
 \hat{H}_{\text{mw}}=g_{ij}\left(s_{ij}e^{iw_{ij}t}\ket{i}\bra{j}+h.c.\right).
 \label{Driving_Hamiltonian}
 \end{equation}
 Here $\omega_{ij}=E_i-E_j$ is the energy splitting between levels $i$ and $j$, $s_{ij}$ is the matrix element, describing the strength of the corresponding microwave transition, $g_{ij}$ is the amplitude of the microwave pulse, proportional to the magnetic field amplitude. In the simple picture described above, that neglects all nonsecular terms, one calculates $w_{ij}$ as the difference between eigenvalues of the Hamiltonian (\ref{hyperfine_Hamiltonian}), that includes only the secular terms
 \begin{equation}
 \begin{aligned}
	\hat{H}&=\left(-|\Omega|+\sum_{j=\{x,y,z\}}h_j^-\hat{I}_j\right)\ket{-}\bra{-},\\
	&+\left(|\Omega|+\sum_{j=\{x,y,z\}}h_j^+\hat{I}_j\right)\ket{+}\bra{+},\\
	&+\left(2D_{gs}+\sum_{j=\{x,y,z\}}h_j^1\hat{I}_j\right)\ket{1}\bra{1}.
 \end{aligned}
 \label{simple_Hamiltonian}
 \end{equation}
 To parametrize $h_j^e$, $e\in\{+,-,1\}$, we introduce $\theta^{\pm}, \phi^{\pm}, \theta^1, \phi^1$ according to
 \begin{eqnarray}
 \begin{aligned}
    \boldsymbol{h^{e}}&=\left(h^e_x, h^e_y, h^e_z\right)\\
    &=|h^e|\left(\sin\theta^e\cos{\phi^e},\sin\theta^e\sin{\phi^e},\cos\theta^e\right),\\
 \end{aligned}
 \end{eqnarray}
 for $e\in\{+,-,1\}$.
 
 The eigenstates of the system in Fig. \ref{fig:levels} will take the form
 \begin{eqnarray}
 \begin{aligned}
	 &\psi^e_p=\ket{e}\begin{pmatrix}
	 \cos(\frac{\theta^e}{2}) \\
	 \sin(\frac{\theta^e}{2})e^{i\phi^e}\\
	 \end{pmatrix},\\
	 &\psi^e_m=\ket{e}\begin{pmatrix}
	 \sin(\frac{\theta^e}{2}) \\
	 -\cos(\frac{\theta^e}{2})e^{i\phi^e}\\
	 \end{pmatrix},\\
	 \label{explicit states}
 \end{aligned}
 \end{eqnarray}
with the eigenvalues being $E_{1,2}=-|\Omega|\pm|h^-|/2$, $E_{3,4}=|\Omega|\pm|h^+|/2$, $E_{5,6}=2D_{gs}\pm|h^1|/2$. The state with index $p$ corresponds to the upper state of the hyperfine doublet, with index $m$-to the lower one. If the system is driven with a microwave field pointing in the y-direction, $s_{ij}$ from equation (\ref{Driving_Hamiltonian}) can be explicitly calculated using the states (\ref{explicit states}). For example $s_{26}$ will take the form
  \begin{eqnarray}
  \begin{aligned}
  s_{26}&=\braket{\psi^-_p|S_y|\psi^1_p}\\
  &=\braket{-|S_y|1}\begin{bmatrix}
  \cos(\frac{\theta^-}{2})\cos(\frac{\theta^1}{2})+\sin(\frac{\theta^-}{2})\sin(\frac{\theta^1}{2}) \\
  \end{bmatrix}.\\
  \end{aligned}
  \end{eqnarray}
 The last factor here comes from the scalar product of two nuclear spin wave functions and its absolute value is $\cos\left(\alpha\right)=(\boldsymbol{h^-}\cdot \boldsymbol{h^1})/(|\boldsymbol{h^-}||\boldsymbol{h^1}|)$.

To gain universal control over the system we propose to use eight microwave pulses, that couple the levels $1,2,3,4$ to the levels $5,6$ (Fig.\ref{fig:levels}). The control Hamiltonian will then contain eight copies of (\ref{Driving_Hamiltonian}) with $i\in\{5,6\}$, $j\in\{1,2,3,4\}$. If we change into a rotating frame, in which all the six levels have the same energy, the control Hamiltonian will take the form
\begin{eqnarray}
\hat{H}_{\text{mw}}=\sum_{i\in\{5,6\},j\in\{1,2,3,4\}}g_{ij}s_{ij}\ket{i}\bra{j}+\mathrm{h.c.}.
\end{eqnarray}
Adjusting the two amplitudes $g_{61}$, $g_{62}$, we can couple any superposition of levels $\ket{1}$, $\ket{2}$ to the state $\ket{6}$. Let us now choose any pair of orthogonal nuclear spin states $\ket{0_n}$, $\ket{1_n}$, then
\begin{eqnarray}
\begin{aligned}
&\ket{-,0_n}=\alpha\ket{1}+\beta\ket{2},\\
&\ket{-,1_n}=-\beta^*\ket{1}+\alpha^*\ket{2}.\\
\end{aligned}
\end{eqnarray}

\begin{figure}[h]
	\includegraphics[width=0.48\textwidth]{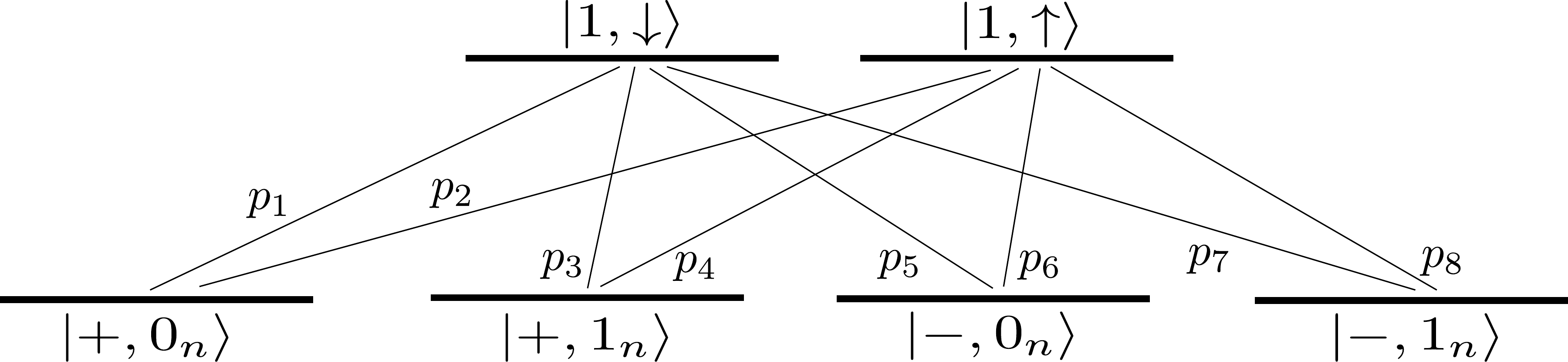}
	\caption{Eight different pulse protocols to couple any basis state of a two-qubit register to the upper electronic state of the NV center.}
	\label{fig: r.f.levels}
\end{figure}

If we now apply the two pulses simultaneously, one coupling the level $1$ to the level $6$ and one coupling the level $2$ to $6$, such that $g_{61}=g\alpha^*/s_{61}$, $g_{62}=g\beta^*/s_{62}$, the control Hamiltonian will take the form $\hat{H}_{\text{mw}}=g\ket{6}\bra{-,0_n}+\mathrm{h.c.}$. Analogously, if we define $g_{61}=-g\beta/s_{61}$, $g_{62}=g\alpha/s_{62}$, we will obtain the control Hamiltonian $\hat{H}_{\text{mw}}=g\ket{6}\bra{-,1_n}+\mathrm{h.c.}$. These pairs of pulses are the new control pulse protocols that can couple $\ket{-,0_n}$ and $\ket{-,1_n}$ to the level $\ket{6}$. Similarly, we can define in total eight new control pulse protocols that will couple the levels $\ket{-,0_n}$, $\ket{-,1_n}$, $\ket{+,0_n}$, $\ket{+,1_n}$ to the levels $\ket{5}$ and $\ket{6}$. We name these new pulse protocols $p_1,p_2,...., p_8$ and show them in the Fig. \ref{fig: r.f.levels}.
 Each of these pulses has a magnitude $|p_i|$ and a phase $f_i$. In the rotating frame all states have the same energy, but in a real system the possibility to apply each of the given pulse protocols separately is based on the fact that the energy differences between levels $1,2,3,4$ and the levels $5,6$ in Fig. \ref{fig:levels} have eight different values and the corresponding transitions can be resolved. In Fig. \ref{fig:levels} the transition from level $1$ to level $5$ and from level $2$ to level $6$ are closest to each other. Choosing the parameters as discussed in Appendix \ref{appB}, we numerically diagonalize the Hamiltonian ($\ref{hyperfine_Hamiltonian}$) and find that the closest resonance frequencies differ by $36$ MHz. The inverse of this value is $28$ ns and it sets the limit to how fast we can operate our system. If we apply one frequency tone close to a given resonance, we want its effect on other transitions to be negligible. This is only possible if the coupling amplitude is a lot smaller than $36$ MHz, that means a gate operation time should be a lot larger than $28$ ns. If the gate time is too low, nonresonant transitions will start to affect the gate fidelity. Nevertheless, this still leaves enough room for a submicrosecond control of the two-qubit register.  

In this section we based our description on the simplified Hamiltonian (\ref{simple_Hamiltonian}), that neglects the nonsecular interaction of the electron spin with nonparallel magnetic field and nonsecular hyperfine interaction terms. The same arguments can be given if one uses a more rigorous effective Hamiltonian (\ref{appA:approximate_hyperfine_Hamiltonian}), that we derive in the Appendix \ref{appA}. This Hamiltonian takes into account the nonsecular interaction terms and is valid to second order perturbation theory.
\section{Initialization and Readout}
 \label{Sec3}
\begin{figure}[h]
	\includegraphics[width=0.4\textwidth]{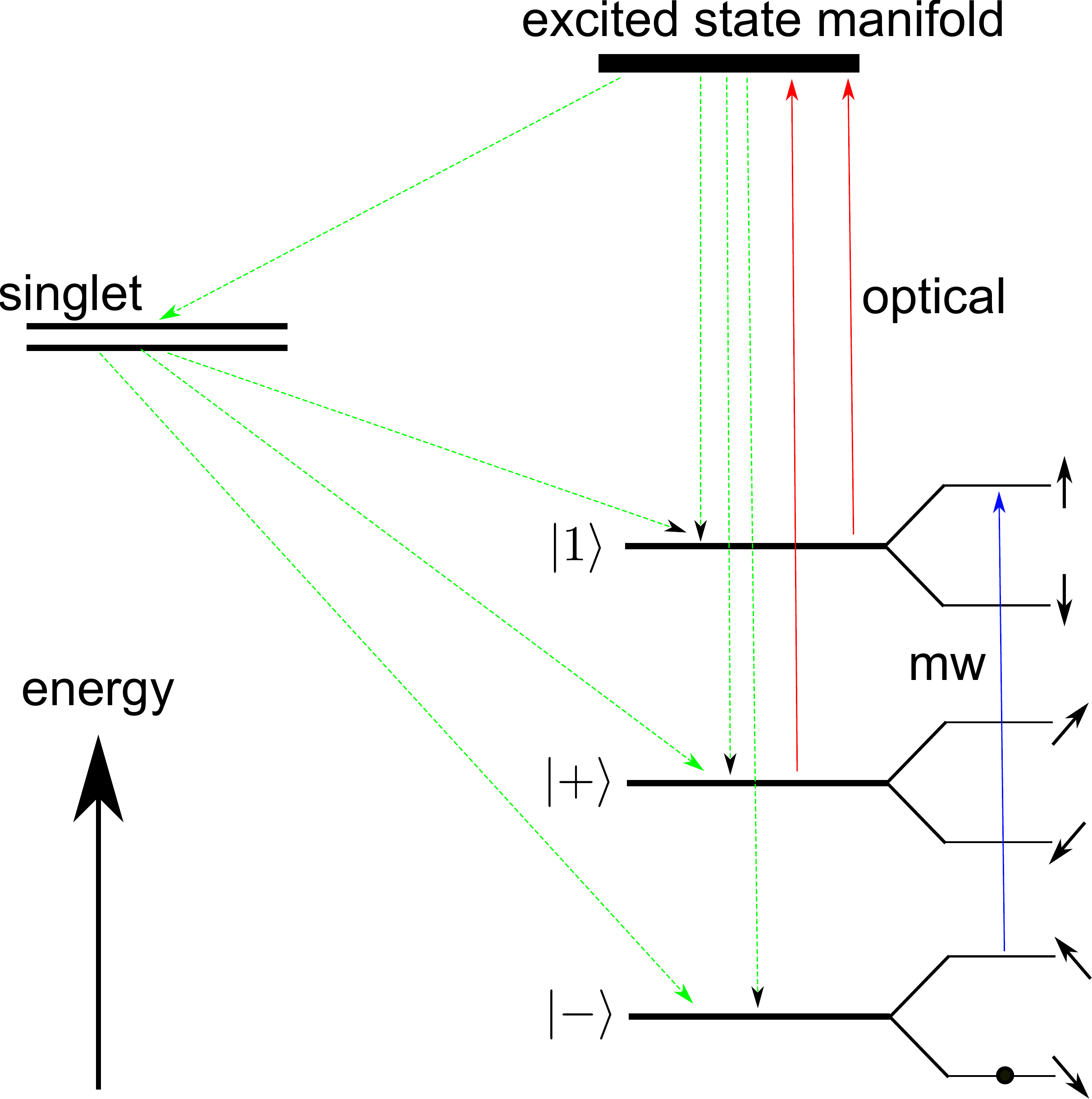}
	\caption{State initialization protocols. Optical pulses (red) pump the electronic states $\ket{1}$ and $\ket{+}$ to the NV excited state manifold. An additional simultaneous microwave tone (blue arrow) pumps one of the two lowest hyperfine sublevels. Green arrows illustrate incoherent mechanisms that return the population to the ground state. After many excitation cycles the system becomes trapped in the lower hyperfine level (black dot). }
	\label{fig: Initialization scheme}
\end{figure}

In this section we show how to perform initialization and readout of the system in the regime, suggested in the previous section, when $B_z=D_{gs}/\gamma_e$ and $|\Omega|\gg||A||$. Our proposal to perform initialization and readout of the system is based on coherent population trapping (CPT) \cite{RevModPhys.77.633} and resembles the scheme used in Ref. \cite{Robledo2011}. Figure \ref{fig: Initialization scheme} shows the procedure.
 Here the excited state manifold consists of an orbital doublet, spin triplet and a hyperfine doublet, thus forming a twelve-dimensional space. The exact Hamiltonian, governing the dynamics of the excited state manifold will be given in the Appendix \ref{appB}. The electronic levels $\ket{1}$, $\ket{+}$ are coupled to the excited state manifold through optical excitation, shown as red arrows in the Fig. \ref{fig: Initialization scheme}. The frequencies of the optical fields are such that the level $\ket{-}$ is out of resonance, while the  other two levels $\ket{1}$, $\ket{+}$ are coupled close to resonance. In Appendix \ref{appB} we show that one can achieve this with a single frequency optical field in the relevant magnetic field regime. In order to initialize the system in the lowest level of Fig. \ref{fig: Initialization scheme}, an additional microwave pulse is required, shown as a blue arrow in Fig. \ref{fig: Initialization scheme}. Whenever the lowest level is not populated, it will be brought to the excited state manifold through a combination of microwave and optical pulses. From there, the population will incoherently decay back to the ground state manifold through the channels marked with green arrows in the Fig. \ref{fig: Initialization scheme}. Then the process repeats itself until after many cycles of optical and microwave excitation the population becomes trapped in the lowest level of Fig. \ref{fig: Initialization scheme}. We performed a numerical simulation of this initialization procedure and showed that in 100 $\mu \rm{s}$ the system can be initialized with a fidelity of $98\%$, in agreement with the results obtained for a similar procedure in \cite{Robledo2011}. The details of the simulation and the relevant parameters are given in Appendix \ref{appB}.

Read out can be performed in a similar manner. Let us assume that we want to know whether the system is in a state $\psi$. We first perform a gate that takes $\psi$ to the lowest level of Fig. $\ref{fig: Initialization scheme}$, followed by the initialization procedure. The absence of luminescence intensity indicates the system was initially in the state $\psi$, the presence of luminescence intensity indicates the opposite measurement result.

\section{Universal set of holonomic gates}
\label{Sec4}
\begin{figure}[b]
	\includegraphics[width=0.3\textwidth]{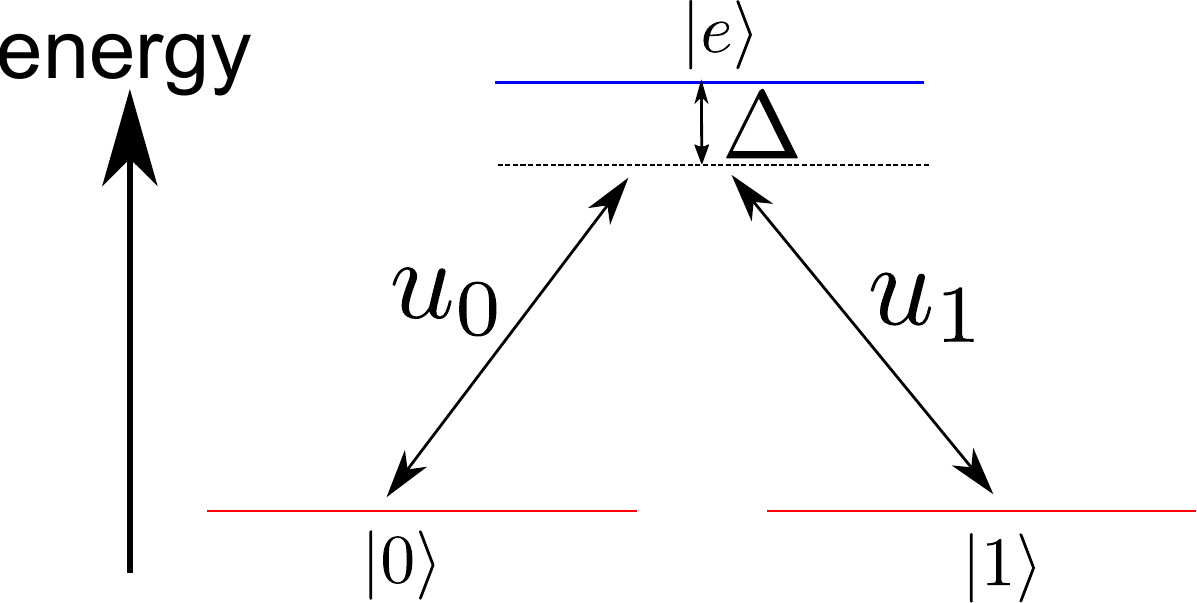}
	\caption{Energy levels in $\Lambda$-configuration. Two different laser fields with interaction matrix elements $u_0$ and $u_1$ (controlled by the laser amplitudes and phases) and a common detuning $\Delta$ couple the ground states $\ket{0}$ and $\ket{1}$ to the same excited state $\ket{e}$. This generates a nontrivial unitary operation on the lower levels, when the system is driven to the $\ket{e}$ state and back.}
	\label{fig: lambda system}
\end{figure}

 Our proposal is to preform nonabelian holonomic gates \cite{SJOQVIST201665} using the setup described in the previous sections. Each such gate involves three levels, two ground states ($\ket{0}$ and $\ket{1}$), that are treated as a logical basis (or part of the logical basis) and one excited state $\ket{e}$, as shown in Figure~\ref{fig: lambda system}. The lower states are coupled to the excited state with two different coupling amplitudes $u_0$ and $u_1$, at frequencies that are both detuned from resonance by the same value $\Delta$, as shown in Fig.~\ref{fig: lambda system}. In this scheme we assume that the ground state levels have the same energy, but this is not necessary if the rotating wave approximation is valid, which holds for the transitions between the NV center levels in the regime we consider here. In that case one can always describe the system in the rotating frame, where the ground states have the same energy. Going to a rotating frame modifies the laser frequency, such that they would have the same detuning with respect to the excited state. This immediately suggests that the validity of our scheme relies on how well one can control the frequency in experiment and how good the phase and frequency locking between the lasers can be made.  Modern technology allows very precise frequency standards and very good laser locking, that is why we assume this is not an issue that can affect the fidelity of our gates and from now on we will assume the rotating wave picture and precise coupling frequencies and phases. 

\begin{figure}[t]
	\begin{center}
		\includegraphics[width=0.5\textwidth]{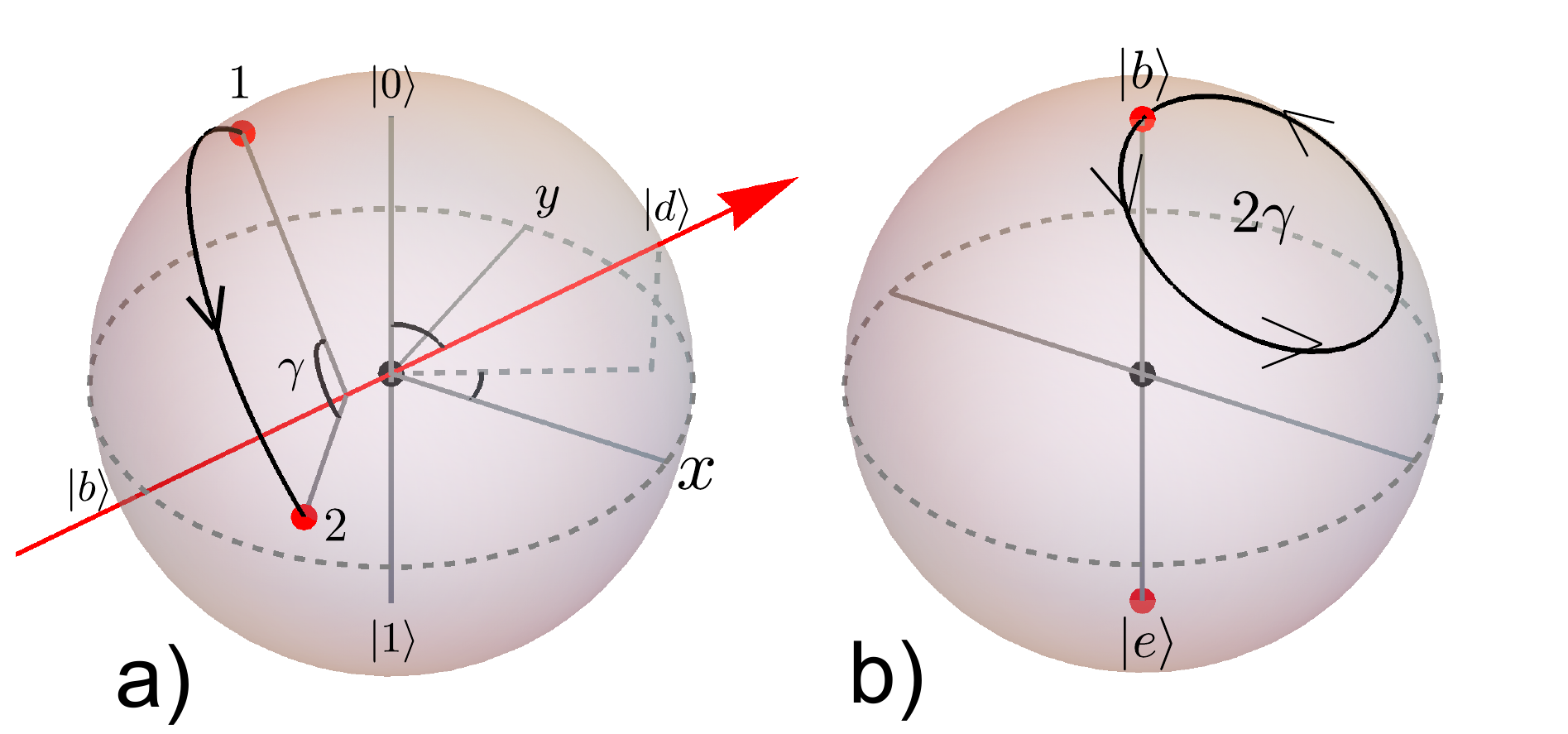}
		\caption{Geometrix phase. (a) A rotation by $\gamma$ on the Bloch sphere, spanned by $\ket{0}$ and $\ket{1}$, corresponding to the transformation $\alpha\ket{d}+\beta\ket{b}\rightarrow\alpha\ket{d}+e^{i\gamma}\beta\ket{b}$. 
		(b) The angle $\gamma$ has a purely geometric origin and equals to half of the solid angle traced by the cyclic trajectory of $\ket{b}$ on the Bloch sphere spanned by $\ket{b}$ and $\ket{e}$.}
		\label{fig:BlochSpheres_rotations}
	\end{center}
\end{figure}

In the rotating frame of the couplings, the $\Lambda$-system Hamiltonian reads
\begin{equation}
\hat{H}=(u_0\ket{0}\bra{e}+u_1\ket{1}\bra{e}+h.c.)+\Delta\ket{e}\bra{e}.
\label{eq:Lambda_Hamiltonian}
\end{equation}
We can introduce the bright ($\ket{b}$) and dark ($\ket{d}$) states according to
\begin{equation}
\begin{aligned}
&\ket{b}=\frac{u_0}{u}\ket{0}+\frac{u_1}{u}\ket{1},\\
&\ket{d}=\frac{u_1^*}{u}\ket{0}-\frac{u_0^*}{u}\ket{1},
\end{aligned}
\end{equation}
where $u=\sqrt{|u_0|^2+|u_1|^2}$.
In this new basis the Hamiltonian takes the following form
\begin{equation}
\hat{H}=u(\ket{b}\bra{e}+\ket{e}\bra{b})+\Delta\ket{e}\bra{e},
\label{eq:bright_state_Hamiltonian}
\end{equation}
Assuming $u_0$ and $u_1$ are time independent (the coupling pulses are flat), this Hamiltonian generates the following evolution of the state $\ket{b}$
\begin{equation}
\ket{b}\rightarrow e^{-i\Delta t/2}\big[\ket{b}\cos(\omega t)-i\bigg(\frac{u}{\omega}\ket{e}-\frac{\Delta}{2\omega}\ket{b}\bigg)\sin(\omega t)\big],
\label{cyclicity1}
\end{equation}
where we introduced $\omega=\sqrt{u^2+\Delta^2/4}$. If the couplings are only switched on for a time $t=\pi/\omega$, the state $\ket{b}$ will evolve into
\begin{equation}
\ket{b}\rightarrow e^{i\gamma}\ket{b},\text{ }	\gamma=\pi-\frac{\pi\Delta}{\sqrt{\Delta^2+4u^2}}.
\label{cyclicity2}
\end{equation}
If we start with a superposition $\alpha\ket{d}+\beta\ket{b}$, this transformation generates the following gate
\begin{equation}
\alpha\ket{d}+\beta\ket{b}\rightarrow\alpha\ket{d}+e^{i\gamma}\beta\ket{b},\label{state}
\end{equation}
that is equivalent to a rotation by the angle $\gamma$ about the axis, connecting $\ket{b}$ and $\ket{d}$ on the Bloch sphere (Fig.~\ref{fig:BlochSpheres_rotations} a). It is worth mentioning that the Hamiltonian (\ref{eq:bright_state_Hamiltonian}) gives zero expectation value of the energy for any state in the ground state space. When the state (\ref{state}) evolves in time, the energy expectation value remains zero because the Hamiltonian is time independent. The angle $\gamma$ thus has purely geometric nature and on a Bloch sphere spanned with $\ket{b}$ and $\ket{e}$ equals half of the solid angle traced by the cyclic trajectory of $\ket{b}$ (Fig. \ref{fig:BlochSpheres_rotations} b).
Since the bright state and the angle $\gamma$ can be chosen arbitrarily, any unitary can be generated in the space of $\ket{0}$ and $\ket{1}$. Indeed, the axis of rotation on the Bloch sphere can be controlled by the relative strength and phase of the couplings $u_0$ and $u_1$, while the angle of rotation is controlled with the detuning $\Delta$ and can be tuned to take any value from zero to $2\pi$. What needs to be taken care about is the timing, meaning that the couplings $u_0$ and $u_1$ must be switched on exactly for the time $\pi/\omega$.

We now show how to construct a universal set of gates to control the two-qubit register in the magnetic field regime given in section \ref{Sec2}, when $B_z=D_{gs}/\gamma_e$ and $|\Omega|\gg||A||$. Universal control requires that one can generate each of the eight  microwave tones $p_1, p_2,....., p_8$, without driving any other transitions.
 Figure \ref{fig: r.f.levels} suggests different ways to identify a $\Lambda$-system, discussed before. Using the pulses $p_1$ and $p_3$ we can create a $\Lambda$-system that allows one to perform universal gates on the nuclear spin controlled by the state of an electron spin. More precisely, the nuclear spin state is flipped only if the electron spin is in the state $\ket{+}$. Analogously, using the pulses $p_5$ and $p_7$ one arrives at gates on the nuclear spin, controlled with the $\ket{-}$ state of the electron spin. Performing the same gate first using the pulses $p_1$ and $p_3$ and then the pulses $p_5$ and $p_7$, one performs universal single qubit gates on the nuclear spin. In exactly the same way, gates on the electron spin controlled with the state of a nuclear spin can be performed combining pulses $p_1$ with $p_5$ and $p_3$ with $p_7$. Thus, universal holonomic computation with the two-qubit register can be achieved. 
 
 For example, the CPHASE gate can be performed if the control pulse protocol $p_8$ is switched on with zero detuning for the time $\tau=\pi/|p_8|$. That is equivalent to switching two laser pulses that resonantly couple the levels $1$ and $2$ to the level $6$ (Fig.~\ref{fig:levels}). The amplitudes and phases of the lasers are adjusted such that the microwave Hamiltonian takes the form 
 \begin{eqnarray}
 \begin{aligned}
 \hat{H}_{\text{mw}}&=\ket{6}\left(g_{61}s_{61}\bra{1}+g_{62}s_{62}\bra{2}\right)+\mathrm{h.c.}\\
 &=p_8\ket{6}\bra{-,1_n}+\mathrm{h.c.}
 \end{aligned}
 \end{eqnarray}
 
 In addition to being able to generate a universal set of gates, one must also make sure that the generated gates are of sufficiently high fidelity to be useful for quantum information applications. There are two qualitatively different sources of gate errors in our scheme. The first one is Markovian noise, related to $T_1$ and $T_2$ times of the NV center and the nuclear spin. In principle, we could include this noise into our consideration in a similar way as we treated spontaneous emission to simulate the initialization process, by means of a Lindblad equation. On the other hand, considering that our gates are rather fast (we assume each of the two couplings in a Lambda system to be $2.5$ MHz, which yields the gate operation time of approximately $285$ ns), and the $T_1$ and $T_2$ times can be made hundreds of microseconds for NV centers, we will not consider these processes as the main source of errors and neglect them in our consideration of the gate robustness. The second source of errors comes from the non-Markovian environment of the NV center. For example, the $^{13}C$ nuclear spins create a random magnetic field which remains constant during the gate operation, i.e. field noise having a correlation time much larger than the gate time. The most relevant effect of this random magnetic fields is that the energies of the electronic levels in Figure~\ref{fig:levels} become ill defined, as the field interacts with the electronic spin. During the gate operation this leads to fluctuations of the detuning $\Delta$ as well as to the emergence of some random energy difference between $\ket{0}$ and $\ket{1}$ in Figure~\ref{fig: lambda system} or between $\ket{+}$ and $\ket{-}$ in Figure~\ref{fig:levels}.  In principle, these two effects are present at the same time, but for simplicity, we will consider them separately. This will also allow us to see which of the two effects is more important. In Ref.~\cite{Rong2015} it was shown that the detuning $\Delta$ between the electronic levels $\ket{0}$ and $\ket{1}$ has a Gaussian distribution with $\sigma=0.131$ MHz. In our scheme we can then assume this value to be $1.5$ times bigger, as the levels $\ket{\pm}$ have a contribution of the $\ket{-1}$ electronic level, which causes them to go in the other direction in energy by half the value with respect to the level $\ket{1}$. The energy fluctuation between $\ket{+}$ and $\ket{-}$ is caused by the random magnetic field, perpendicular to the symmetry axis of the NV. We assume this value to be $\sqrt{2}\sigma$, as according to the definition of $|\Omega|$ above $2\delta|\Omega|=\sqrt{2}\gamma_e\delta B_\perp$, and so the standard deviation of this value is $\sqrt{2}$ times bigger than that of $\gamma_e B_z$, if we assume all directions of the random field to be equivalent. Another effect reducing the gate fidelity consists in the fluctuations of microwave coupling amplitudes used to drive the quantum gates. Although we need two different frequency tones to drive our gates, we assume that they arise from the same pulse generator, and so the fluctuations in driving strengths are correlated. In Ref.~\cite{Rong2015} it was shown that these fluctuations obey a Lorenzian distribution with $\gamma=0.0024$ MHz, that we will also assume for our simulations. 
 
 We describe the effect of these imperfections on the gate performance in the following way. Let us assume we are interested in the effect of the detuning fluctuations  ($\delta\Delta$). First we assume that fluctuations are absent and define the pulse sequence that will yield the targeted gate $U$. We then apply this pulse sequence to the initial state of the system ($\rho_{\text{in}}$) many times, assuming different values of the detuning $\Delta$, and obtaining the final density matrix  $\rho_{\text{f}}(\Delta)$ as a function of $\Delta$. We now average this density matrix over $\Delta$ with the distribution function $f(\Delta)$, which yields the action of a quantum channel on our initial state $\rho_{\text{in}}$:
 \begin{equation}
 	\rho_{\text{in}}\rightarrow\mathcal{E}(\rho_{\text{in}})=\int \rho_{\text{f}}(\Delta)f(\Delta)d\Delta.
 \end{equation}
 With this procedure in mind, we are able to compute the average gate fidelity according to
 \begin{equation}
 	F=\int d\psi\braket{\psi|U^{\dagger}\mathcal{E}(\ket{\psi}\bra{\psi})U|\psi}.
 \end{equation}
 As it was shown by Horodecki \cite{Horodecki1999}, this fidelity can be related to the so called entanglement gate fidelity
 \begin{equation}
 F_e=\braket{\phi|U^{\dagger}\openone\otimes\mathcal{E}(\ket{\phi}\bra{\phi})U|\phi},
 \end{equation}
 where $\ket{\phi}$ is some maximally entangled state of the system and ancilla, which has the same dimension as the system. In our case the system of interest is four-dimensional ($d=4$), as we are manipulating the state of two qubits. Thus we need to add a four dimensional ancilla. However, we use two additional levels to perform holonomic gates and thus there might be leakage errors, when the system can be found outside of the logical space with nonzero probability. In this context we note that the relation between average gate fidelity and entanglement fidelity was initially derived for trace-preserving gate operations, i.e., in the absence of leakage errors. For our purpose, the corresponding relation thus needs to be modified and takes the form (see Appendix~\ref{appC})
 \begin{equation}
  F=\frac{dF_e+\mathrm{Tr}(\mathcal{E}(\openone/d))}{d+1},
  \label{eq:avg_gate_fiedlity}
  \end{equation}
  where $d$ is the dimension of the Hilbert space of the system and $\openone/d$ denotes the totally mixed state of the system. Thus, we numerically apply targeted gates through the corresponding maps  $\mathcal{E}$ on two states, once on the maximally entangled state of the system and ancilla to compute the entanglement fidelity and once on the maximally mixed state of the system to compute $\mathrm{Tr}(\mathcal{E}(\openone/d))$. In the absence of leakage errors the latter experession is equal to one and we recover the standard result from Horodecki and Nielsen \cite{Horodecki1999,NIELSEN2002}. 
    
 In Ref.~\cite{Rong2015} optimal control theory was used to construct special gate pulses to reduce the effect of fluctuations of detuning and coupling amplitudes such that the error contributions quadratic in $\sigma$ and $\gamma$ could be eliminated. In contrast, as it was shown in Ref.~\cite{Thomas2011,Johansson2012}, the error of the nonadiabatic holonomic gate is quadratic in detuning and pulse amplitude fluctuation strengths. But this does not necessarily mean that these gates have low fidelity. One has to take into account that the effect of errors for nonadiabatic holonomic gates is qualitatively different from that of their dynamical counterparts \cite{Zheng2016}. Indeed, the detuning error ($\delta\Delta$) and the pulse strength error ($\delta u$) on the one hand modify the expression for $\omega$ in Equation~(\ref{cyclicity1}), thus preventing the cyclic evolution of the state $\ket{b}$. On the other hand, these errors affect the area traced by the trajectory of $\ket{b}$ in Figure~\ref{fig:BlochSpheres_rotations}b, thus modifying the phase $\gamma$ in Equation~(\ref{cyclicity2}). The energy fluctuation between $\ket{+}$ and $\ket{-}$ is yet another type of error that dephases these two levels during the operation of the gate. We consider the effect of these errors on two different gates, CNOT on the electron spin, controlled by the nuclear spin, and CNOT on the nuclear spin, controlled by the electron spin. Ideally, without errors, the average gate fidelity is equal to unity. We controlled the precision of our simulation such that the gate fidelity is accurate up to errors of the order $10^{-4}$. We choose the gate time in our simulation to be approximately $285$ ns, obtaining a gate fidelity of $0.998$ limited by  undesired nonresonant transitions caused by each of the laser drives. We find that in the presence of realistic detuning errors $\delta\Delta$, the average gate fidelity for both gates drops to $0.995$. For the error in the coupling strength ($\delta u$), we did not observe any noticeable fidelity decrease within the calculation precision. The random energy fluctuation between $\ket{+}$ and $\ket{-}$ leads to memory errors that degrade the quantum state even when no gates are performed. Still, for the CNOT gate, performed on the electron spin, this error has an additional relevance as $\ket{0}$ and $\ket{1}$ in Figure~$\ref{fig: lambda system}$ are subject to energy fluctuation with respect to each other and the level $\ket{e}$. On the contrary, for the CNOT on the nuclear spin, $\ket{0}$ and $\ket{1}$ would belong to the same electronic level and would thus fluctuate in the same direction, that is equivalent to $\delta\Delta$ error, considered before. We thus consider the effect of random energy fluctuation between $\ket{+}$ and $\ket{-}$ only on the CNOT gate, performed on the electron spin. We find that the fidelity drops to $0.985$. This includes the effect of random energy fluctuation between $\ket{+}$ and $\ket{-}$ during the gate operation time. The infidelity due to this type of noise is found to be $0.013$ including the effect of always-on memory errors. We estimate the overall gate infidelity to be $0.019$, if we assume the infidelity to be additive in this limit, and using an infidelity of $0.003$ originating from $\delta\Delta$ fluctuations. This simplistic estimate provides us with a  lower bound on the gate fidelity in our scheme of around $0.976$ whereas the theoretical gate fidelity of the CNOT gate for the pulses generated with optimal control techniques was calculated to be $0.9927$ \cite{Rong2015}. Of course the direct comparison is hard to make, as the setup for the two schemes are very different. For example, in our simulation we assume the gate time to be approximately $285$ ns, compared to  $696$ ns \cite{Rong2015}. Changing the gate time in our scheme leaves some more room for increasing the gate fidelity, but already at this stage our simulations indicate that nonadiabatic holonomic gates in our scheme can be quite robust against certain types of errors, even using simple square pulses, rather than the pulses generated with optimal control theory.

\section{Discussion}
In this work we have shown how to perform universal quantum computing on a two qubit register, consisting of the electron spin of a negatively charged nitrogen-vacancy center in diamond and the nuclear spin of a nearby $^{13}$C atom. Although we only considered the carbon atom of the shell closest to the vacancy due to the strong hyperfine interaction, our method can be extended to control the carbon atoms further away. We estimate that the magnitude of the dipole-dipole hyperfine interaction for $^{13}$C atoms that are twice as far from the vacancy as the closest carbon is such that the transitions $\ket{1}$ to $\ket{5}$ and $\ket{2}$ to $\ket{6}$ can still differ by $1$ MHz and thus the register can be manipulated using microwave transitions only. The density functional theory calculations \cite{PhysRevB.77.155206} that also take into account the Fermi contact term reveal there are approximately $40$ carbon atoms around the vacancy with hyperfine constants greater that $2$ MHz, which suggests that there are more than 3 closest $^{13}$C atoms, to which our method can be applied. It is still possible to perform universal gates on these atoms until the electron spin decoheres, but in that case one also has to include the nitrogen nuclear spin into consideration. Our method can readily be extended to include the nitrogen nuclear spin through the hyperfine interaction Hamiltonian
\begin{equation}
\hat{H}_{N}=A_{||}\hat{S}_z\hat{I}_z+A_{\perp}(\hat{S}_x\hat{I}_x+\hat{S}_y\hat{I}_y),
\end{equation}
with the hyperfine constants being $A_{||}=-2.16$ MHz, $A_{\perp}=-2.6$ MHz \cite{PhysRevB.92.020101}. Using this Hamiltonian, the Knight field acting on the nitrogen nuclear spin can be calculated in the same way as it was done for the closest shell carbon atom.
Other proposals exist to perform universal microwave control on the registers of coupled nuclear and electron spins \cite{Hodges, Khaneja}. They differ from our method in the sense that their gates are not geometric and universality in those schemes requires an external magnetic field acting on the nuclear spins to add up with the hyperfine Knight field and thus create two nonparallel axes of rotation, while in our scheme we only rely on the hyperfine field, that is stronger than the external magnetic field.
Strong hyperfine interaction has its disadvantages in that it decoheres the nuclear spin very fast. Going to a rotating frame picture reveals that the nuclear spin is not affected by the dephasing in the ground state space of the electron spin ($T_2$), but the relaxation processes ($T_1$), as well as reinitialization of the electron spin affect the nuclear spin dramatically \cite{Dutt1312}. Still our scheme can be used to perform universal quantum computation, for example, to gain increased sensitivity of an NV based quantum sensor \cite{Zaiser2016}. We also note that although we only considered our scheme applied to the nuclear spin strongly coupled to the electronic spin, it would also be possible to consider it with respect to weakly coupled nuclear spins. A lot of research is done on the control of these registers \cite{arXiv:1905.02095,PhysRevLett.109.137602,PhysRevLett.123.050401,Taminiau2014} and configuration, in which the levels $\ket{0}$ and $\ket{-1}$ are mixed due to nonparallel magnetic field could offer new pathways to control such registers.

%\clearpage
\appendix

\section{Effective Hamiltonian for hyperfine interaction in each electronic level, valid to second order perturbation theory} 
\label{appA}
In this appendix we treat the coupled nuclear and electron spin system in the presence of nonparallel magnetic field in a rigorous way. We fix $B_z=D_{gs}/\gamma_e$ and introduce the phase factor $\phi$ according to the equation $B_x-iB_y=B_{\perp}e^{i\phi}$. At this field and in this notation one obtains
\begin{equation}
\begin{split}
	\hat{H}_e&=D_{gs}\hat{S}_z^2+\gamma_e\left(B_x\hat{S}_x+B_y\hat{S}_y+B_z\hat{S}_z\right)\\
	&=2D_{gs}\ket{1}\bra{1}+\frac{\gamma_e}{2}B_\perp\left(e^{i\phi}S_{+}+e^{-i\phi}S_{-}\right).\\
\end{split}
\end{equation}
We further introduce $\Omega=\frac{e^{3i\phi/2}}{2}\gamma_eB_{\perp}$ and the new basis states
\begin{equation}
\begin{split}
	&\ket{+}=\frac{1}{\sqrt{2}}\left(e^{i\phi/2}\ket{0}+e^{-i\phi/2}\ket{-1}\right),\\
	&\ket{-}=\frac{1}{\sqrt{2}}\left(e^{i\phi/2}\ket{0}-e^{-i\phi/2}\ket{-1}\right),\\
\end{split}
\end{equation}
so the Hamiltonian $H_e$ takes the form
\begin{equation}
\begin{split}
	&\hat{H}_e=\hat{H}^0+\hat{H}_e^2,\\
	&\hat{H}^0=2D_{gs}\ket{1}\bra{1}+|\Omega|\left(\ket{+}\bra{+}-\ket{-}\bra{-}\right),\\
	&\hat{H}_e^2=\Omega\left(\ket{1}\bra{+}+\ket{1}\bra{-}\right)+\Omega^*\left(\ket{+}\bra{1}+\ket{-}\bra{1}\right).\\
\end{split}
\end{equation}
We now introduce the hyperfine and the nuclear spin Zeeman interactions  $\hat{H}_{\text{hf}}$, $\hat{H}_{\text{n}}$ into the system. We split the hyperfine interaction into secular and nonsecular terms as
\begin{equation}
\begin{split}
	&\hat{H}_{hf}=\hat{H}_{hf}^1+\hat{H}_{hf}^2\\
	&\hat{H}_{hf}^1=\sum_{k=\{1,+,-\}}\ket{k}\bra{k}\sum_{i,j=\{x,y,z\}}A_{ij}\hat{I}_j\braket{k|\hat{S}_i|k},\\
	&\hat{H}_{hf}^2=\sum_{\tilde{k}\neq k}\ket{k}\bra{\tilde{k}}\sum_{i,j=\{x,y,z\}}A_{ij}\hat{I}_j\braket{k|\hat{S}_i|\tilde{k}}.\\
\end{split}
\end{equation}
Our aim now is to obtain an effective Hamiltonian in each of the three electronic subspaces. We achieve this using the formalism of the Schrieffer-Wolf transformation \cite{Winkler:684956,BRAVYI20112793}. The basic idea is to find a basis change that brings to zero non-secular terms $\hat{H}_e$, $\hat{H}_{hf}^2$ up to a certain order of magnitude. In order for this procedure to work we have to assume that $|\Omega|\ll2D_{gs}$, $||A_{ij}||\ll2\Omega$. Following this procedure, we introduce an antihermitian matrix S, that obeys the equation
\begin{equation}
SH_e^0-H_e^0S=-\hat{H}_{hf}^2-\hat{H}_e^2.
\end{equation}
After performing this procedure, we obtain
\begin{equation}
\begin{split} 
	S&=\frac{\Omega}{2D_{gs}-|\Omega|}\ket{1}\bra{+}+h.c.\\
	&+\frac{\Omega}{2D_{gs}+|\Omega|}\ket{1}\bra{-}+h.c.\\
	&-\sum_{k\neq\tilde{k}}\frac{\sum_{ij}A_{ij}\hat{I}_j\braket{k|S_{i}|\tilde{k}}}{W_{\tilde{k}}-W_k},
\end{split}
\end{equation}
where $W_{1}$, $W_{+}$ and $W_{-}$ are $2D_{gs}$, $|\Omega|$ and $-|\Omega|$ respectively.
Now we can calculate the effective Hamiltonian in each of the three electronic subspaces
\begin{equation}
	\hat{H}_{\mathrm{eff}}=\hat{H}_e^0+\hat{H}_{n}+\hat{H}_{hf}^1+\frac{1}{2}\left[S,\hat{H}_{hf}^2+\hat{H}_e^2\right].
\end{equation}
 From this expression it follows that the interaction of the electron spin with the transverse magnetic field leads to the renormalization of the energies of the bare electronic states $D_{gs}$, $|\Omega|$, $-|\Omega|$ to the new values 
 \begin{equation}
 \begin{split} 
 &\tilde{D}_{gs}=D_{gs}\left(1+\frac{2|\Omega|^2}{4D_{gs}^2-|\Omega|^2}\right)\\
 &\Omega_{\pm}=|\Omega|\left(1\mp\frac{|\Omega|}{2D_{gs}\mp|\Omega|}\right)\\
 \end{split}
 \end{equation}
Let us also introduce the corrections to hyperfine terms due to the interaction of electron spin with the transverse magnetic field
 \begin{equation}
 \begin{split} 
 &C_{\pm}=\frac{2\mathrm{Re}\left[\Omega\braket{\pm|S_i|1}\right]}{2D_{gs}\mp|\Omega|}\\
 \end{split}
 \end{equation}
For the effective Hamiltonian we then obtain
 \begin{equation}
 \begin{split} 
 &\hat{H}_{\mathrm{eff}}=2\tilde{D}_{gs}\ket{1}\bra{1}+\Omega_{+}\ket{+}\bra{+}-\Omega_{-}\ket{-}\bra{-}\\
 &+\sum_{ij}A_{ij}\hat{I}_j\Bigl(\braket{1|S_i|1}+C_{+}+C_{-}\Bigr)\ket{1}\bra{1}\\
 &+\sum_{ij}A_{ij}\hat{I}_j\Bigl(\braket{+|S_i|+}-C_{+}\Bigr)\ket{+}\bra{+}\\
 &+\sum_{ij}A_{ij}\hat{I}_j\Bigl(\braket{-|S_i|-}-C_{-}\Bigr)\ket{-}\bra{-}\\
 &+\frac{1}{2}\sum_{ij}\frac{A_{ij}A_{\tilde{i}\tilde{j}}}{2|\Omega|}\braket{+|S_i|-}\braket{-|S_{\tilde{i}}|+}\hat{I}_j\hat{I}_{\tilde{j}}\ket{+}\bra{+}\\
 &-\frac{1}{2}\sum_{ij}\frac{A_{ij}A_{\tilde{i}\tilde{j}}}{2|\Omega|}\braket{-|S_i|+}\braket{+|S_{\tilde{i}}|-}\hat{I}_j\hat{I}_{\tilde{j}}\ket{-}\bra{-}\\
 &+\hat{H_n}.\\
 \end{split}
 \label{appA:approximate_hyperfine_Hamiltonian}
 \end{equation}
  The first line here arises due to the interaction of the electron spin with magnetic field. Lines two to four describe the secular terms of the hyperfine interaction corrected due to nonsecular terms of the electron spin interaction with the magnetic field. Lines five and six take into account nonsecular part of the hyperfine interaction, that mixes the electron spin states $\ket{+}$, $\ket{-}$. We neglect the effect of this interactions that mixes $\ket{+}$, $\ket{-}$ with $\ket{1}$, because it is small compared to all other effects.
  
  \section{Details of the dynamics simulation} 
  \label{appB}
  In this appendix we give details of the simulation we performed to model the initialization fidelity. The ground state of the NV is an orbital singlet, spin triplet and we describe it with the Hamiltonian (\ref{hyperfine_Hamiltonian}). The excited state is treated as an orbital doublet, spin triplet and we describe it with the following Hamiltonian \cite{DOHERTY20131} 
  \begin{equation}
  \begin{split} 
 \hat{H}_{es}
 &=g_{es}^{||}\mu_BB_z\hat{S}_z+2\mu_B\left(B_x\hat{S}_x+B_y\hat{S}_y\right)\\
 &-\lambda\sigma_y\hat{S}_z+l\mu_BB_z\sigma_y\\
 &+D_{zz}\left(\hat{S}_z^2-\frac{1}{3}S(S+1)\right)\\
 &+D_{xy}\left(\sigma_z(\hat{S}_y^2-\hat{S}_x^2)-\sigma_x\{\hat{S}_x,\hat{S}_y\}\right)\\
 &+D_{xz}\left(\sigma_z\{\hat{S}_x,\hat{S}_z\}-\sigma_x\{\hat{S}_y,\hat{S}_z\}\right)\\
 &+\sum_{i,j=\{x,y,z\}}A_{ij}^{es}\hat{S}_i\hat{I}_j.\\
 &+\gamma_n\boldsymbol{B}\cdot\boldsymbol{\hat{I}}\\
  \end{split}
  \label{appB:excited state Hamiltonian}
  \end{equation}
  
  Here $\sigma_x$, $\sigma_y$, $\sigma_z$ are Pauli matrices that act in the basis of $\ket{E_x}$, $\ket{E_y}$ of the orbital doublet. The first line describes the Zeeman interaction, the second line takes into account spin-orbit coupling and the interaction of magnetic field with the orbital angular momentum $\hat{L}$ (described with an operator $\sigma_y$ in the relevant subspace). Lines three, four and five describe the spin-spin interaction. The sixth line takes into account the hyperfine interaction in the excited state. The seventh line gives the Zeeman interaction for the nuclear spin. The strength of the corresponding interactions is taken from Ref. \cite{Bassett1333} and is listed in Table \ref{tab: parameters}. The strength of the hyperfine interaction is given in the basis when z-axis coincides with the direction from the vacancy to the $^{13}C$ atom. In this basis the hyperfine tensor is diagonal, with the biggest eigenvalue ($A_{||}^{es}$) corresponding to the vector along z-direction. The excited orbital states can decay to the ground state through a spin-conserving photon emission with the rate $\Gamma_{ge}$ each. The corresponding decay operators are $O_x=\ket{A_2}\bra{E_x}$ and $O_y=\ket{A_2}\bra{E_y}$, where $\ket{A_2}$ is the ground state orbital singlet. The excited state $\ket{A_1}$, that is the eigenstate of the spin-orbit and spin-spin part of the Hamiltonian (\ref{appB:excited state Hamiltonian}), can also decay to the singlet level $\ket{s}$ with the rate $\Gamma_{se}$. The corresponding decay operator is $O_3=\ket{s}\bra{A_1}$. The singlet state can decay to the ground through three channels, to the state with spin $0$ at the rate $\Gamma_{0s}$ and corresponding decay operator $O_4=\ket{A_2,m_s=0}\bra{s}$ or to the states with spins $\pm1$ with the rate $\Gamma_{pms}$ and corresponding decay operators $O_5=\ket{A_2,m_s=+1}\bra{s}$ and $O_6=\ket{A_2,m_s=-1}\bra{s}$ . The values of the decay rates are taken from Ref. \cite{Bassett1333} and are listed in Table \ref{tab: parameters}. We assumed the z-component of the magnetic field to be $D_{gs}/\gamma_e$. The transverse magnetic field is assumed to point in the x-direction and have the value of $500$ G. We show that at this magnetic field one optical field is enough to couple the electronic levels $\ket{+}$ and $\ket{1}$ to the excited state $\ket{E_x}$ (Figure \ref{fig: Initialization scheme}), while leaving $\ket{-1}$ out of resonance. The microwave magnetic field pulse in Fig. \ref{fig: Initialization scheme} is assumed to point in the y-direction. The corresponding optical and microwave Rabi frequencies $\Omega_{o}$ and $\Omega_{mw}$ respectively are given in Table \ref{tab: parameters}. The optical and microwave driving Hamiltonians are given by
    \begin{equation}
    \begin{split} 
    &\hat{H}_{o}=\Omega_o(\ket{E_x}\bra{A_2}e^{i\omega_1 t}+\ket{A_2}\bra{E_x}e^{-i\omega_1 t}),\\
    &\hat{H}_{mw}=\Omega_{mw}\hat{S}_y\sin(\omega_2 t).\\
    \end{split}
    \label{appB: driving Hamiltonians1}
    \end{equation}
  
  We solve the Lindblad equation
  \begin{equation}
  \begin{split} 
  \dot{\rho}=&-\frac{i}{\hbar} [ \hat{H}_{gs}+ \hat{H}_{es}+ \hat{H}_{o}+ \hat{H}_{mw},\rho ]\\
  &+\sum_{i=1}^{6}\Gamma_i\left(\hat{O}_i\rho\hat{O}_i^\dagger-\frac{1}{2}\hat{O}_i^\dagger\hat{O}_i\rho-\frac{1}{2}\rho\hat{O}_i^\dagger\hat{O}_i\right),
  \end{split}
  \label{appB: driving Hamiltonians2}
  \end{equation}
  assuming the duration of the optical and microwave pulses of $100$ $\mu s$. Our simulation reveals that assuming the system to be initially in equal superposition of the six ground states, after such procedure the system will be trapped in the lowest ground state with the probability $97 \%$.
  \begin{table}[h!]
  	\begin{center}
  		\caption{The values used to simulate the initialization fidelity.}
  		\label{tab: parameters}
  		\begin{tabular}{|l|r|}
  			\hline
  			\bf{Parameter}  & \bf{Value} \\
  			\hline
  			g-factor $g_{es}^{||}$  & $2.15$\\
  			\hline
  			spin-orbit constant $\lambda$  & $5.33$ GHz\\
  			\hline
  			l   & $0.1$\\
  			\hline
  			axial spin-spin constant $\mathrm{D_{zz}}$  & $1.44$ GHz \\
  			\hline
  			transverse spin-spin constant $\mathrm{D_{xy}}$  & $1.54/2$ GHz \\
  			 \hline
  			 transverse spin-spin constant $\mathrm{D_{xz}}$   & $154/\sqrt{2}$ MHz \\
  			  \hline
  			  $A_{||}^{es}$  & $126$ MHz \\
  			   \hline
  			    $A_{\perp}^{es}$  & $56.7$ MHz \\
  			    \hline
  			    $\gamma_n$  & $0.001$ MHz/G \\
  			    \hline
  			    $\Gamma_{ge}$ & $83.3$ MHz \\
  			    \hline
  			    $\Gamma_{se}$ & $400$ MHz \\
  			    \hline
  			    $\Gamma_{0s}$  & $1.5$ MHz \\
  			    \hline
  			    $\Gamma_{pms}$  & $0.58$ MHz \\
  			    \hline
  			     $\Omega_{o}$  & $25$ MHz \\
  			     \hline
  			      $\Omega_{mw}$  & $10$ MHz \\
  			      \hline
  		\end{tabular}
  	\end{center}
  \end{table}
   %The parameters of the simulation are optical orbital coupling, microwave coupling. We model the system's ground state as a spin triplet, orbital singlet. The excited state is modelled as orbital doublet, pin triplet. We also include intermediate level to model the singlets. The system has a spontaneous orbital decay channel from the excited states to the ground state. It also has nonradiative decay path from the excited state to the singlet level and from the singlet to the ground state with spin dependent decay rates.  We obtain the fidelity of initialization of 98\%, that agrees with the results obtained for a similar procedure in \cite{Robledo2011}. 

   \section{Average gate fidelity for trace-non-preserving maps}\label{appC}
   In this appendix we give a proof of equation \eqref{eq:avg_gate_fiedlity}, extending the result of Horodecki \textit{et al.} \cite{Horodecki1999} and Nielsen \cite{NIELSEN2002}. In their work they relate the average gate fidelity of a quantum map to its entanglement fidelity. 
   
   For convenience, let us briefly recall some definitions necessary to comprehend equation \eqref{eq:avg_gate_fiedlity}. A quantum map is a $\mathbb{C}$-linear map which takes density operators, acting on some state Hilbert space, to another density operator. Especially, $\mathcal{E}$ preserves hermiticity and is completely positive and trace-preserving. The average fidelity is defined as 
   \begin{align}
	   F(\mathcal{E})\equiv\int\bra{\psi}\mathcal{E}(\psi)\ket{\psi}\mathrm{d}\psi,
	   \label{eq:avg_fid}
   \end{align} 
   where $\mathcal{E}(\psi) \equiv \mathcal{E}(\ket{\psi}\bra{\psi})$.
   The average gate fidelity $F(\mathcal{U}\circ\mathcal{E})$ (where $\mathcal{U}(\rho)\equiv U^\dagger\rho U$) serves as a measure of how well a desired unitary gate $U$ is approximated by $\mathcal{E}$. The integration measure $\mathrm{d}\psi$ is taken to be the Haar measure, $\int\mathrm{d}\psi=1$ and is integrated over the state space of the system. The entanglement fidelity, as introduced in \cite{Schumacher1996}, requires an extension of the system Hilbert space $\mathcal{H}$ to a larger Hilbert space $\tilde{\mathcal{H}}=\mathcal{Q}\otimes\mathcal{H}$ where $\mathcal{Q}$ is a copy of $\mathcal{H}$. The entanglement fidelity of $\mathcal{E}$ is then given by
   \begin{align}
	   F_e(\mathcal{E})\equiv\bra{\phi}(\openone\otimes\mathcal{E})(\phi)\ket{\phi},
	   \label{eq:ent_fid}
   \end{align}
   where $\ket{\phi}\in\tilde{\mathcal{H}}$ denotes a maximally entangled state. %In equation \eqref{eq:avg_fid} and \eqref{eq:ent_fid} we used $\psi$ %(respectively $\phi$) to denote both, $\ket{\psi}$ and %$\ket{\psi}\bra{\psi}$. 
   Note that, while computing $F(\mathcal{E})$ is generally very hard due to the integration, the entanglement fidelity is much easier to calculate (upon choosing a unitary operator basis). It is thus desirable to have a relation between the two quantities which was found in \cite{Horodecki1999} to be
   \begin{align}
	   F=\frac{d F_e+1}{d+1},
	   \label{eq:horodecki}
   \end{align}
   $d$ being the dimension of the system Hilbert space $\mathcal{H}$.
   
   However, as noted in the main text, if one is interested merely in the evolution of a subspace $\mathcal{H}_s\subset\mathcal{H}$ it can occur that the respective quantum map on that subspace is not trace-preserving and thus violates one of the conditions leading to equation \eqref{eq:horodecki}. In such a case it is desirable to find a generalised relation applying to trace-non-preserving quantum maps as well.
   \theoremstyle{plain}
   \newtheorem*{Claim}{Claim}
   \begin{Claim}
   	Let $\mathcal{E}$ be a trace-non-preserving quantum map acting on density matrices of a $d$-dimensional Hilbert space $\mathcal{H}$. Then its average gate fidelity is given by
   	\[F=\frac{dF_e+\mathrm{Tr}(\mathcal{E}({\normalfont\openone}/d))}{d+1}.\] 
   \end{Claim}
   The general idea of the proof follows the one of Nielsen \cite{NIELSEN2002}, suitably modified for the trace-non-preserving case. We thus state some of the initial steps without showing them explicitly (instead referring to Nielsen's paper) and are more careful with the modified parts.
   \begin{proof}
   		First, we define the twirled map
   		\begin{align}
	   		\mathcal{E}_T(\rho)\equiv\int U^\dagger\mathcal{E}(U\rho U^\dagger)U\mathrm{d}U,
   		\end{align}
   		where $\mathrm{d}U$ denotes the uniform Haar measure on the space of unitary operators $U$ and $\rho$ a density operator, both acting on the Hilbert space $\mathcal{H}$. Note that $\mathcal{E}_T$ inherits linearity from $\mathcal{E}$. It is shown in \cite{NIELSEN2002} that the twirling of $\mathcal{E}$ leaves $F$ and $F_e$ invariant (note that no use is made of the trace-preserving property by Nielsen at this point). Hence, it suffices to prove the relation for $\mathcal{E}_T$. It can be shown that for any unitary $V$
   		\begin{align}
	   		V\mathcal{E}_T(\rho)V^\dagger=\mathcal{E}_T(V\rho V^\dagger).
	   		\label{eq:trafo_condition}
   		\end{align}
   		This holds true for arbitrary density operators $\rho$. Specifically, consider a one-dimensional projector $P$ and its orthogonal complement $Q=\openone-P$ on $\mathcal{H}$. Choose $V$ to be block-diagonal with respect to these subspaces, i.e. $VPV^\dagger=P$. From equation \eqref{eq:trafo_condition} it follows that $V\mathcal{E}_T(P)V^\dagger=\mathcal{E}_T(P)$. Thus, $\mathcal{E}_T(P)$ is block-diagonal as well and can be written in the form 
   		\begin{align}
	   		\mathcal{E}_T(P)=\alpha P+\beta Q=\beta\openone+(\alpha-\beta)P.
	   		\label{eq:depol_channel}
   		\end{align}
   		Note that this is similar to a depolarising channel, however, its trace is not unity in general (this is where the proof deviated from \cite{NIELSEN2002}).
   		The coefficients $\alpha,\beta\in\mathbb{C}$ are determined by the probability of depolarisation and the trace of $\mathcal{E}_T(P)$. First, we can express $\alpha$ as
   		\begin{align}
	   		\alpha=\mathrm{Tr}\mathcal{E}_T(P)-\beta(d-1)
   		\end{align}
   		by taking the trace of equation \eqref{eq:depol_channel}. Note that, since $\mathrm{Tr}\mathcal{E}_T(P)\in\mathbb{R}$, in fact we have $\alpha,\beta\in\mathbb{R}$. Define $\beta\equiv p/d$ and $c\equiv\mathrm{Tr}\mathcal{E}_T(P)$ such that the twirled map can be expressed as
   		\begin{align}
	   		\mathcal{E}_T(P)=\frac{p}{d}\,\openone+(c-p)P.
	   		\label{eq:depol_channel2}
   		\end{align}
   		In the next step we show that, in fact, $p$ and $c$ are independent of the choice of $P$. Note that an arbitrary one-dimensional projector on $\mathcal{H}$ can be obtained as $P^\prime=VPV^\dagger$ with a unitary $V$. Now, applying equation \eqref{eq:trafo_condition} to \eqref{eq:depol_channel2} gives
   		\begin{align}
	   		\mathcal{E}_T(P^\prime)=\mathcal{E}_T(VPV^\dagger)=\frac{p}{d}\,\openone+(c-p)P^\prime.
   		\end{align}
   		This shows that $p$ and $c$ are independent of the choice of $P$ (for $c$ this is apparent from $\mathrm{Tr}(V\mathcal{E}_T(P)V^\dagger)=\mathrm{Tr}\mathcal{E}_T(P)$ as well). Since an arbitrary density operator $\rho$ can be expressed as a linear combination of one-dimensional projectors and $\mathcal{E}_T$ is a linear map, we have
   		\begin{align}
	   		\mathcal{E}_T(\rho)=\frac{p}{d}\,\openone+(c-p)\rho.
   		\end{align}
   		With $\mathcal{E}_T$ of the above form, $F$ and $F_e$ can be calculated explicitly: $F=p/d+c-p$ and $F_e=p/d^2+c-p$. This immediately yields equation \eqref{eq:horodecki} where one in the numerator is replaced by $c$. Finally, since $c$ is independent of the density operator that $\mathcal{E}_T$ acts on, we can express it in a natural way as $c=\mathrm{Tr}\mathcal{E_T}(\openone/d)=\mathrm{Tr}\mathcal{E}(\openone/d)$, where the second equality follows from the definition of $\mathcal{E}_T$ and linearity of the trace. Our formula now follows from the invariance of $F$ and $F_e$ under twirling.
   \end{proof}
	Note that for trace-preserving $\mathcal{E}$ we have $c=1$ and recover equation \eqref{eq:horodecki} from our formula.

  % \newpage

\bibliography{literatureCompile}
\end{document}